\documentclass[a4paper,11pt]{article}
\usepackage{jinstpub} 
\usepackage{lineno}
\usepackage[utf8]{inputenc}
\usepackage{gensymb}

\title{Infrared narrow band emitting quantum dots for high energy physics, medicine and space applications}

\author[a]{T. Choudhury}
\affiliation[a]{Institute of Engineering and Management (IEM)}
\emailAdd{tribikram.choudhury@cern.ch}

\author[b]{Y. Haddad}
\affiliation[b]{Northeastern University}
\emailAdd{yacine.haddad@cern.ch}

\author[c]{M. Doser}
\affiliation[c]{CERN}
\emailAdd{michael.doser@cern.ch}

\abstract{
Infrared quantum dots, operating in the near-infrared (NIR, 700-1400 nm), short-wavelength infrared (SWIR, 1400-3000 nm), mid-infrared (MIR, 3000-8000 nm) and long-wavelength infrared (LWIR, 8000-15000 nm) regions, have promising potential in optoelectronics, nanotechnology and military surveillance applications. The properties of infrared quantum dots exhibit quantum confinement effects, unlike bulk semiconductors, where their bandgap energy and emission wavelength can be precisely tuned by controlling particle size, composition, and surface chemistry. The wide tunability and unique quantum confinement effects in these infrared-emitting materials also make them attractive for both fundamental research, health and space technology. This paper focuses on the synthesis, fabrication and characterisation of polymer-based infrared quantum dots and explores the possible applications of infrared quantum dots in high-energy physics, medicine and astrophysics.

}


\begin{document}
\maketitle
\flushbottom

\section{Introduction}
Infrared (IR) Quantum Dots (QDs), which have emission and absorption beyond the UV-vis electromagnetic spectrum region, are semiconductor nanocrystals that exhibit size-tunable optical and electronic properties that are very interesting for telecommunications, infrared photodetectors, solar cells, and night-vision systems.\cite{Kershaw2013,Ma2010,Sargent2005, Xue2024-ot,Seo2024-gc,Bertram2018,Pradhan2019}. The development of IR QDs relies on advanced colloidal synthesis techniques \cite{Cossairt2016,Semonin2012} which allow precise control over the size, composition, and surface properties of QDs, which are critical for achieving desired optical and electronic characteristics \cite{Hines2003,Hickey2008}. Notable materials used for IR QDs include lead chalcogenides (PbS, PbSe), III-V semiconductors (InAs, InSb), and emerging perovskite-based nanocrystals \cite{Jellicoe2016,Liu2017,Protesescu2017,Hazarika2018}. IR QDs have been extensively studied for their use in optoelectronic devices, such as photodetectors, light-emitting diodes (LEDs), and solar cells. Their tunable band gaps enable efficient light absorption and emission, which is particularly useful for telecommunications and night vision technologies \cite{Supran2015}. The ability of IR QDs to absorb light in the IR spectrum also makes them excellent candidates for next-generation solar cells. These QDs are being integrated into multijunction solar cells to extend the absorption range and improve energy conversion efficiencies \cite{Nozik2010}. Additionally, multiple-exciton generation (MEG) in QDs has the potential to surpass the Shockley-Queisser efficiency limit of conventional photovoltaic materials \cite{Octavi2012}. Recent research includes the introduction of core-shell structures, such as PbSe/CdSe or InAs/CdSe with optimised photoluminescence quantum yields, also with the aim to mitigate environmental degradation \cite{Zhang2014,Kim2015AirStableAE,Morris2017TowardIS,Pietryga2008,Cao2000,Aharoni2006-sm}. The experimental approach of producing 
Polymer-based infrared-emitting quantum dots having promising optical properties, along with the application perspectives of IR QDs in various technological sectors, will be presented in the next sections.

\section{Experimental Results and Discussions}

\subsection{Experimental Technique}
In this experimental work, the synthesis of the IR QDs involves colloidal synthesis of a mixture of nanoparticles doped with trivalent rare-earth ions, followed by dispersion in a toluene solution, mixing with PDMS polymer. Rare-earth ions play a pivotal role in enhancing the performance of optical detectors, laser systems, display technologies, and telecommunication applications. Trivalent rare-earth (RE) ions doped in semiconductor QDs show excellent luminescence properties because of the intra-configurational transitions of 4f electrons shielding well $5s$ and $5p$ electrons \cite{Trac2021,R2019}. Researchers have shown RE ions doping in QDs for use in different applications such as optical converters (down and up-conversion), solid lasers, and optical amplifiers  \cite{Chu2021}. In recent years, down-conversion materials (such as RE ion-doped QDs) have been investigated in many studies because of their potential to increase solar cell efficiency and protection \cite{Mallick2023}. The chemicals used for the synthesis and fabrication of the infrared quantum dots are Erbium (III) nitrate pentahydrate, Zirconyl chloride octahydrate, Propionic acid, Isopropanol, Magnesium nitrate hexahydrate, Ammonium dihydrogen phosphate, Heptane, Toluene and Sylgard PDMS, purchased from Sigma Aldrich. The prepared QDs were spin-coated onto a $30\times 20\rm mm^2$ quartz silica substrate at $3000$ rpm revolution speed for 40 seconds, followed by UV exposure and a baking step at $100^{\circ}~\rm C$ for $12$ min. The fabricated thin film having quantum dots has been denoted here as IR QD. The surface morphology of the IR QDs has been characterised by Field Emission Scanning Electron Microscopy (FESEM), \textsc{SUPRA 55 VP-4132 CARL ZEISS}, X-ray Diffraction (XRD) measurement with a Rigaku X-ray Diffractometer. The transmittance measurement of IR QDs has been carried out by a PerkinElmer Lambda $1050$ UV-visible-NIR spectrometer. The FLS1000 photoluminescence spectrometer with a semiconductor laser diode emitting at 980 nm has been used to measure photoluminescence emission spectra and lifetimes of the IR QDs in the NIR ($1400$ to $1700~\rm nm$).

\subsection{Morphological and Structural Properties}
\subsubsection{Field Emission Scanning Electron Microscopy (FESEM)}

\begin{figure}[h!]
    \centering
    \includegraphics[width=0.45\textwidth]{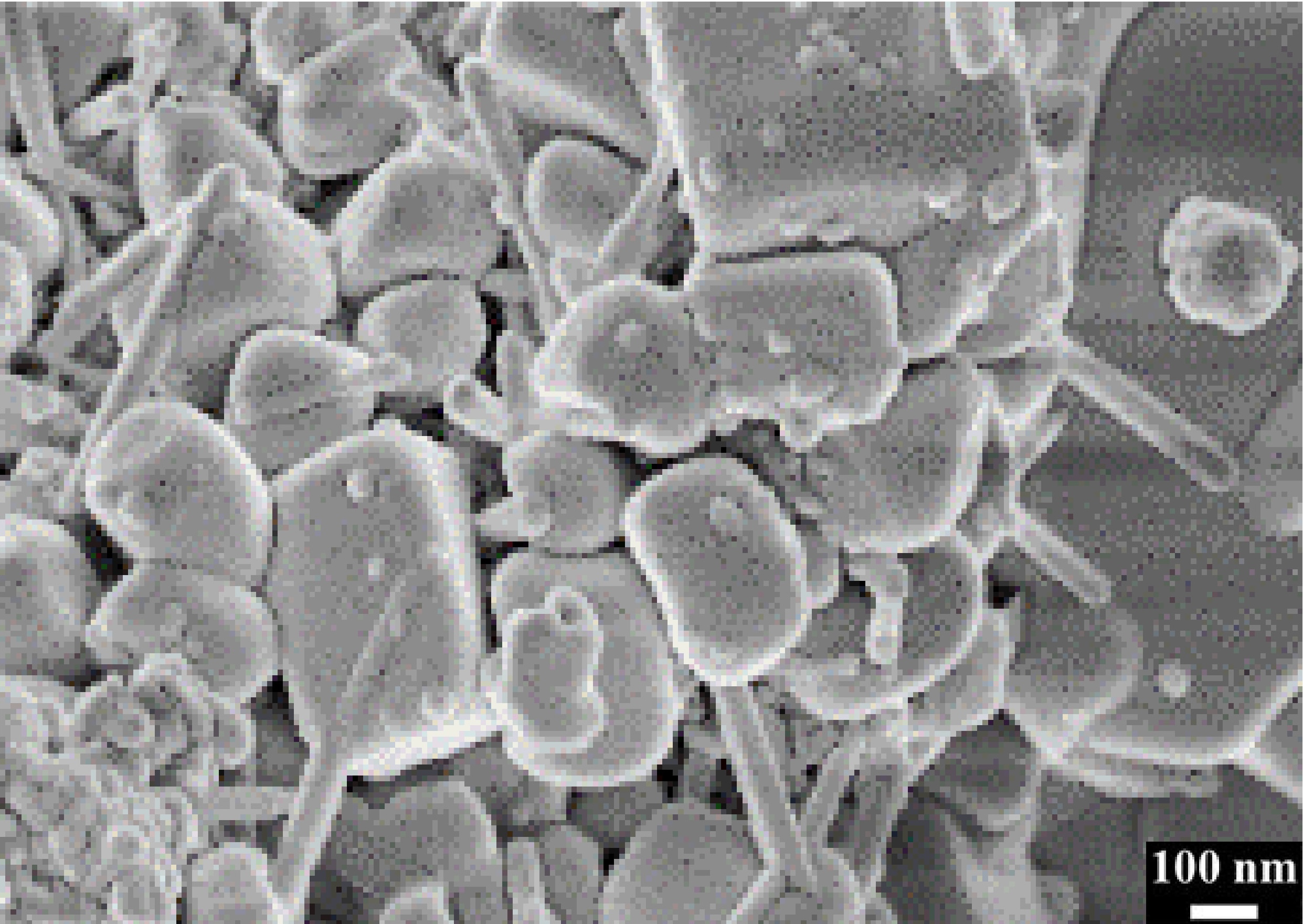}
    \includegraphics[width=0.48\textwidth]{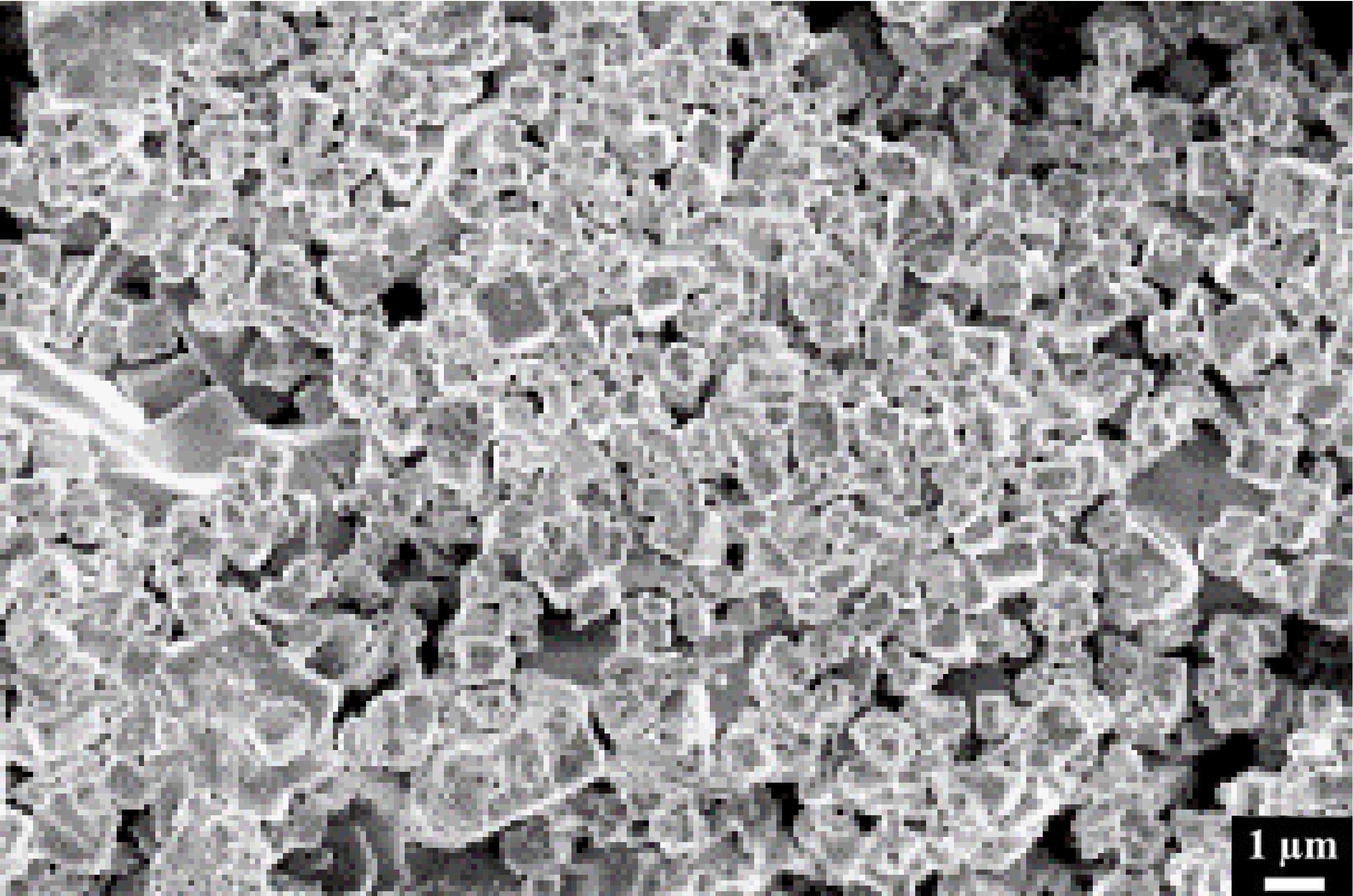}
    \caption{FESEM images of IR QDs}
    \label{fig:FESEM}
\end{figure}

Figure~\ref{fig:FESEM} shows FESEM images of IR QDs, which reveal that the nanoparticles exhibit a smooth surface with minimal roughness or defects with no apparent pores, cracks, or irregularities, indicating good crystallinity. The QD nanoparticles are clear with an average diameter of 11.8 nm, and are predominantly rhombohedral with few spherical or quasi-spherical rounded contours. While most particles are uniform, a few show slight elongation or irregularity, possibly due to coalescence during synthesis or drying \cite{Cassidy2020-ap}. There is negligible aggregation, suggesting soft agglomeration, which is common in nanoparticles due to high surface energy and van der Waals attractions \cite{Talapin2010-xn,Kagan1996-wk,Kovalenko2015-lx,Peng1997-gr}. No large-scale sintering or fusion is observed.

\subsubsection{X-ray Diffraction (XRD) Measurements}

\begin{figure}[h!]
    \centering
    \includegraphics[width=0.8\textwidth]{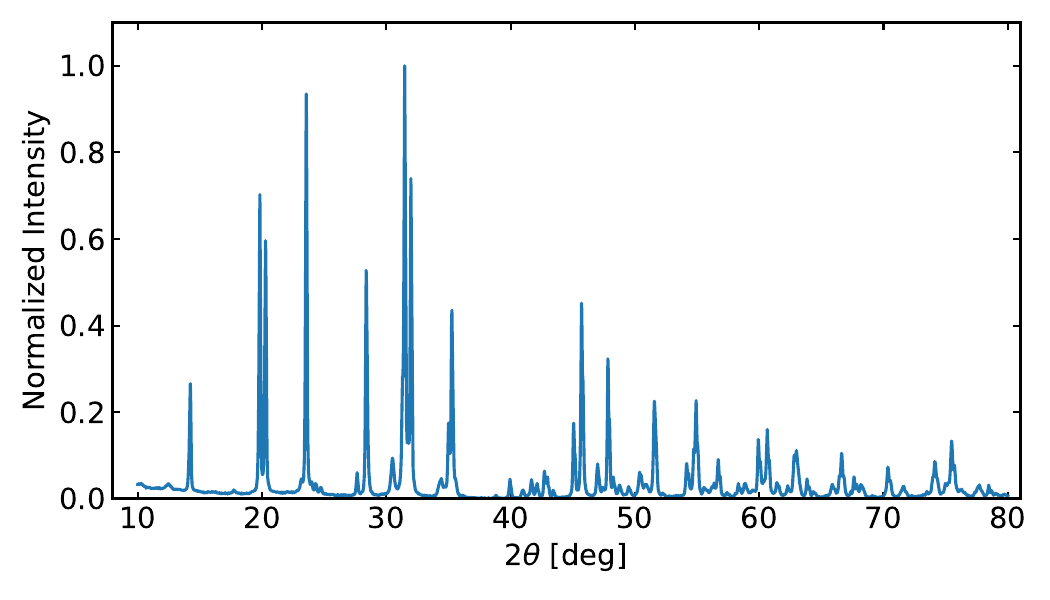}
    \caption{XRD patterns of IR QDs calcined at 900 $^{\circ}$C for 3h}
    \label{fig:XRD}
\end{figure}

XRD analysis was performed for the IR QDs as shown in Figure \ref{fig:XRD} with respective diffraction angle, FWHM, d-spacing, Miller index (hkl) values, respectively, within $2\theta$ (scan range 10 - 80  [deg]) at room temperature. The XRD pattern matches the standard data available in the International Centre for Diffraction Data (ICDD) card number 04-005-5490 with rhombohedral crystal structure and with no additional impurity phase. The XRD pattern has several diffraction peaks at 2$\theta$ and the corresponding reflection from crystallographic planes or Miller index (hkl) are shown in Table~\ref{tab:2theta-table}. The dominant diffraction peak observed around 31.6° is accompanied by a shoulder near 33.0°, suggesting either overlapping reflections from closely spaced crystallographic planes or contributions from multiple phases \cite{Alivisatos1996-ym,Murray1993-gr}. The corresponding interplanar spacings (d) are calculated for the mentioned (hkl) values in Table~\ref{tab:2theta-table} using Bragg’s law, n$\lambda$ = 2d sin$\theta$, where $\lambda$ = 0.15406 nm is the X-ray wavelength, $\theta$ is the diffraction angle in the XRD pattern, and 'n' is the order of reflection (n=1). The sharpness of the peaks indicates a relatively well-crystallised material with no amorphous content, matching well with the observed and computer-calculated analysis from the FESEM measurement. However, the clustering of reflections in the low-angle region suggests possible preferred orientation or texture effects \cite{Niederberger2006-cq}. The variation in the diffraction patterns is due to the lattice strain for the presence of trivalent Erbium ions in the quantum dots ~\cite{Bindu2014-db,Ayvacikli2012-ax,Zhang2015-lk,Gobechiya2008,Komuro2000,Lebour2009,Nair2015HighlyEC}.

\begin{table}[]
    \centering
    \begin{tabular}{c|c|c}
         \hline
            \textbf{2$\theta$ (degrees)} & \textbf{Miller Index (hkl)} & \textbf{d  (nm)}\\
            \hline
            14.21633 & (012) & 0.62 \\
            19.81928 & (104) & 0.45 \\
            20.27881 & (110) & 0.44 \\
            23.55775 & (113) & 0.38 \\
            31.60301 & (116) & 0.28 \\
            28.39678 & (024) & 0.32 \\
            35.28081 & (300) & 0.25 \\
            45.73561 & (128) & 0.2 \\
            51.60126 & (137) & 0.18 \\
            54.91399 & (318) & 0.17 \\
            59.92305 & (327) & 0.15 \\
            60.66534 & (146) & 0.15 \\
            60.04894 & (1310) & 0.15 \\
            62.95416 & (2113) & 0.15 \\
            74.14523 & (3114) & 0.13 \\
            75.52016 & (342) & 0.13 \\
            \hline
         
    \end{tabular}
    \caption{2$\theta$ values and corresponding Miller indices (hkl)}
    \label{tab:2theta-table}
\end{table}

\subsection{Optical Properties}

\subsubsection{Photoluminescence (PL) Emission and Transmittance Measurement}

\begin{figure}[h!]
    \centering
    \includegraphics[width=0.7\textwidth]{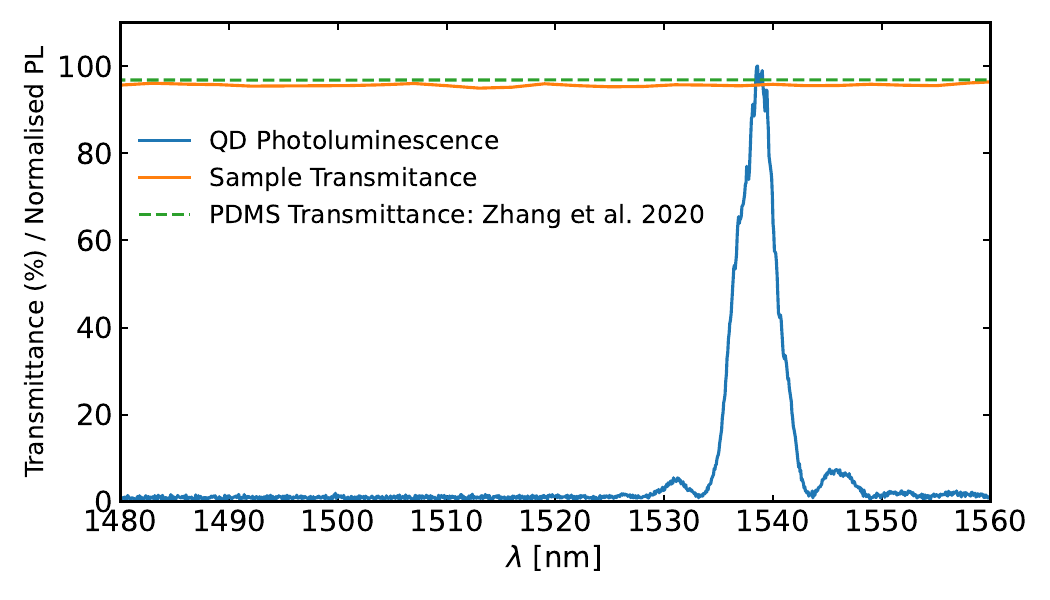}
    \caption{Transmitance spectra of the supporting matrix PDMS sample compared to measurements taken from~\cite{Zhang2020,ZHANG2020107063}, and the measured photoluminescence of the IR-QD.}
    \label{fig:Transmission}
\end{figure}

Figure \ref{fig:Transmission} illustrates the PL emission spectra of IR QD under a 980 nm excitation of the semiconductor diode laser measured at moderate temperature. The IR QDs have a strong PL emission with a narrow band at 1540 nm with a calculated full-width half-maximum (FWHM) of 24.69 nm. This narrow FWHM indicates that the quantum dots are all of very similar size; these IR QDs are thus highly suitable for optoelectronic applications, having promising best-in-class optical properties. The transmittance spectrum has been investigated over the spectral range from 1400 nm to 1700 nm and is compared with the transmittance of the supporting matrix PDMS sample~\cite{Zhang2020,ZHANG2020107063} (Figure \ref{fig:Transmission}).

\subsubsection{Lifetime Measurement}


\begin{figure}[h!]
    \centering
    \includegraphics[width=0.8\textwidth]{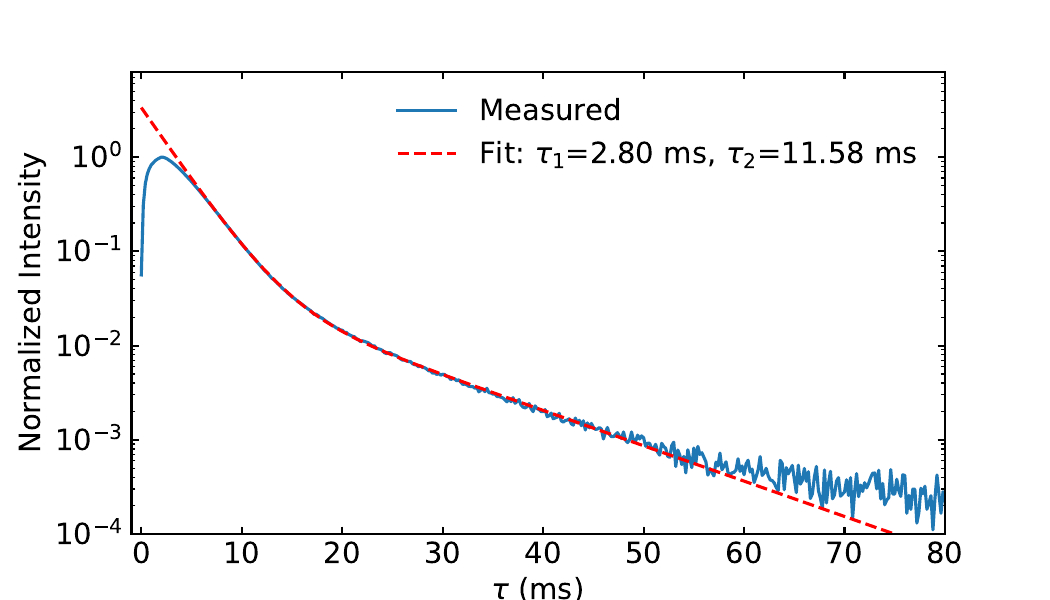}
    \caption{Lifetime of IR QDs}
    \label{fig:Lifetime}
\end{figure}

Figure \ref{fig:Lifetime} shows the PL lifetime of IR QDs measured after exciting with a 980 nm excitation source at room temperature. The PL lifetime of IR QDs was monitored at 1540 nm and then fitted with a single exponential function. The average PL lifetime $\tau_{av}$ was calculated as follows \cite{Kolesnikov2019-wt};

\begin{equation}\label{eq:lifetime}
    \tau_{av} = \frac{Q_1 \tau_1^2 + Q_2 \tau_2^2}{Q_1 \tau_1 + Q_1 \tau_2}
\end{equation}

where $Q_1$ and $Q_2$ are pre-exponential factors, $\tau_1$ and $\tau_2$ denote the rapid and slow lifetimes for the exponents, respectively. The average PL lifetime obtained from equation (2.1) is 3.46 ms for IR QDs.

\section{Future Perspectives for High Energy Physics}
With the experimental results indicating that these IR QDs show promise, their implementation in high-energy physics, specifically in dark matter detection, fixed-target experiments, or collider experiments, can be considered. Recent studies propose leveraging QDs as scintillating targets to detect sub-GeV dark matter through electron interactions. The high photoluminescence quantum yield and tunability of QDs allow for the detection of low-energy events, providing a new avenue for direct dark matter searches.

One of the possible applications of the IR QDs can also be in the area of data transmission. Silicon transparency beyond roughly 1200 nm opens an infrared window that can be explored with high-energy physics detectors. By embedding $1310$ to $1550\,\text{nm}$ optical links directly on tracker modules, modulated light can pass hundreds of microns of silicon with negligible attenuation, reducing copper mass and heat, although coping with multi terabyte per second hit rates anticipated at the HL-LHC and future colliders~\cite{Aberle:2749422} will require faster PL decay times than currently achieved. Radiation-tolerant silicon photonics platforms that integrate Ge on Si photodiodes, Mach Zehnder or electro absorption modulators, and on-chip wavelength division multiplexers have withstood total ionising doses exceeding one Grad and neutron fluences beyond $10^{17}\,\mathrm{n_{eq}\,cm^{-2}}$ in recent irradiation campaigns~\cite{7027245}, foreshadowing lightweight fibre sparse readout architectures.

\begin{figure}[h!]
    \centering
    \includegraphics[width=0.6\textwidth]{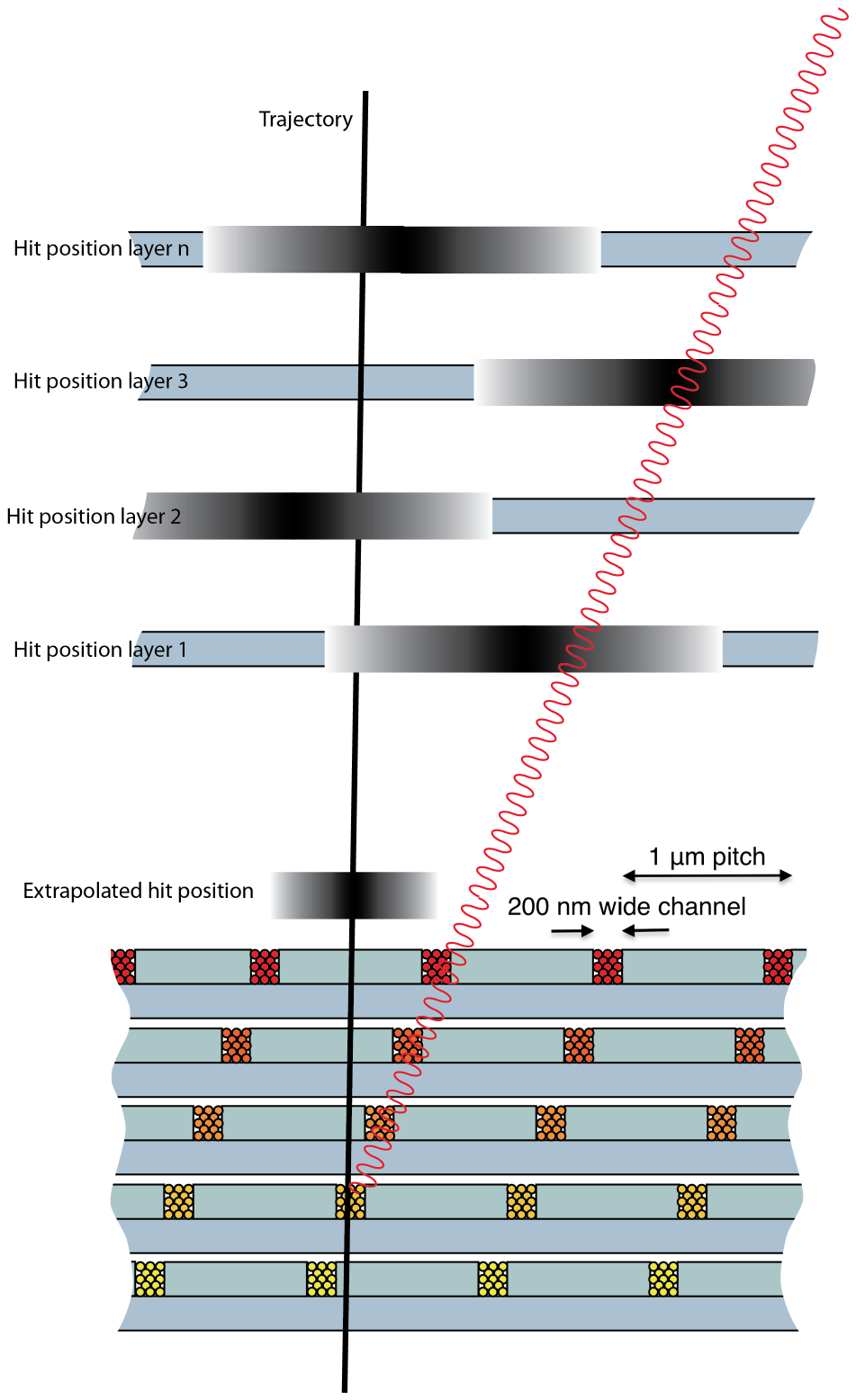}
    \caption{Schematic of a five-layer IR quantum dot chromatic tracker. The track is extrapolated towards the chromatic layer from the multi-layer silicon tracker (the individual and extrapolated hit uncertainties are schematically indicated in black); the detected IR photon then allows uniquely assigning the corresponding disambiguated 200 nm channel to the track. Each layer consists of a substrate (light blue), on top of which a PMMA template (light green) defines the 200 nm wide troughs in which the nanodots are deposited. Prior to assembly, the PMMA is dissolved (not shown).}
    \label{fig:multilayer_detector}
\end{figure}

This same IR transparency of (non fully metalised back plane) silicon detectors can also enable the development of detectors where IR emitting quantum dots are placed closer to the interaction point of collider experiments than the multi-layer vertex tracking detectors typically constructed from multiple layers of silicon pixel and strip detectors. Colloidal nanodots can be deposited with sub-100 nm precision on an O($\mu$m)-thick substrate where the locations of the nanodots are defined through electron beam lithography of a (dissolvable) PMMA template~\cite{Pambudi2024}. As sketched in figure~\ref{fig:multilayer_detector}, individual layers consisting of, e.g. 200 nm wide strips (of quantum dots emitting at a defined layer-dependent wavelength) every 1 micron can thus be constructed. The number of layers and the periodicity are determined by the number of uniquely identifiable nanodot wavelengths, which in turn is influenced by their emission linewidth. If multiple layers are precisely positioned relative to each other (e.g. using ultrasonic linear piezo stages that are precise down to tens of nm) and stacked, a staggered luminescent structure can be assembled, each layer emitting at a different wavelength. If a charged particle traverses this stack, the detection of fluorescence photons produced as the particle traverses a specific group of nanodots allows for the unique assignment of the detected fluorescence photon through its wavelength to a particular quantum dot strip, with potentially 100 nm precision, modulo the overall periodicity. Disambiguation within this overall periodicity (1 micron in this example) is achieved by extrapolating the charged particle trajectory reconstructed using the surrounding silicon micropatterned detectors. Although fluorescence photons can be optically imaged behind the (IR-transparent) silicon layers and their arrival times can be measured, only a limited reconstruction of their origin is possible. The scheme, which still needs to be validated experimentally, is thus limited with respect to the maximal local track density. Different nearby tracks cannot be uniquely assigned a particular photon wavelength, and hence a nanodot position. 

Infrared light is equally powerful for alignment, calibration, and low-threshold calorimetry. Silicon strip and pixel sensors behave as semi-transparent photodiodes, so passing narrow-band infrared lasers through the sensitive volume enables micron-scale deformation monitoring, an approach realised in the CMS laser alignment system that drives forty $1075\,\text{nm}$ beams through $434$ tracker modules ~\cite{duarte2009alignment}. Studies for the International Linear Collider have demonstrated implant layouts and antireflection treatments that raise silicon transmittance to about 50\% near $1080\,\text{nm}$ while preserving charge-collection efficiency. Quantum dot films tuned to telecom wavelengths can add bright monochromatic beacons or up-conversion layers on sensor surfaces. In contrast, noble gaseous and crystalline scintillators possess secondary emission bands between $850$ and $1600\,\text{nm}$ that, when coupled to InGaAs avalanche photodiodes, deliver dual colour timing and sub-electron-volt thresholds for dark matter or axion searches.
Looking forward, a coordinated programme that qualifies silicon photonic components under radiation, refines sensor processing for infrared transparency, and engineers quantum dot coatings that remain robust above $500\ \text{kGy}$ promises to embed infrared functionality throughout future experiments. The expected benefits span massless data links, self-calibrating trackers, and quantum-enhanced rare event detectors, highlighting how advances in telecom and quantum information technology can be repurposed to meet the extreme environments and precision demands of next-generation particle physics.

\section{Future Perspectives for Medicine}
The tunability of the infrared quantum dots in the near-infrared (NIR, 700-1400 nm) and mid-infrared (MIR, 3000-8000 nm) ranges is extraordinarily promising for medical imaging. Infrared-emitting quantum dots are capable of wider and broader access to tagging drugs in infected tissue with higher resolution in comparison to visible quantum dots due to fewer photons getting absorbed and diffused in the infrared region \cite{Cassette2013-hu,Zhao2018NearIQ}. The attractive biocompatibility of the infrared quantum dots is very promising, and experimentally synthesised IR QDs can be a potential candidate for both in vitro and in vivo medical imaging. They can serve as fluorescent markers in biological tissues, with the potential to provide high photoluminescence quantum yields without toxicity, given their absence of heavy metals such as lead, mercury, and cadmium. Surface coating of QDs with peptides or other biomolecules that selectively bind to specific molecular targets, such as lipids in cell membranes, receptors, or other biomolecular markers, is a well-developed area in bioconjugation and targeted imaging, merging nanomaterials with biochemistry \cite{article}. 

Such coatings facilitate advanced biosensing applications, wherein the interaction between the target molecule and the quantum dot modulates its fluorescence characteristics through mechanisms such as Förster Resonance Energy Transfer (FRET) or fluorescence quenching \cite{Cardoso_Dos_Santos2020-pk}. This enables drug delivery tracking \cite{Hanmante2025-wt}, where functional peptides act as targeting ligands for drug-loaded QDs (or QD-containing nanocarriers), and pathogen detection \cite{Xue2022-mr}, where many viruses and bacteria interact with specific lipid components or lipid-binding peptides. The study of membrane biophysics, where peptide-functionalized QDs can be used as nanoscale probes to track the dynamics of lipid rafts \cite{Eggeling2009-ju}, of vesicle fusion, or of membrane curvature changes \cite{Probst2013-mr}.

This technology can be enhanced and optimised further by binding IR-emitting QDs with peptides or biomolecules. This approach shows promise in deep-tissue imaging, targeted sensing \cite{Hong2017-wo}, and theranostics through the combination of targeted peptides with IR-QD fluorescence. Combining drug payload with a lipid-targeting angle of incorporating membrane-insertive peptide sequences to guide IR-QDs into lipid-rich tumour membranes to deliver photothermal or photodynamic therapy agents can also track treatment progress in deep tissue \cite{Uprety2022-qf}. 

IR light has the capability to penetrate biological tissue better than visible light with Near-IR (NIR, 700–1400 nm) and Short-Wave IR (SWIR, 1000–1700 nm) QDs \cite{Wegner2024-ri} having reduced scattering and lower autofluorescence. This can enable non-invasive visualisation of molecular lipid distributions in living animals, which can also be applied to in vivo molecular sensing by targeting specific peptides with IR-QDs bound to a lipid target, followed by localisation of the signal to specific regions via optical readout through tissue. 

Also, IR-QDs are photostable and bright, allowing the tracking of single lipid molecules labelled with QD–peptide conjugates inside a living animal, even through thick tissue layers \cite{Pinaud2010-wl}. Another promising application is in nanodosimetry for particle (proton and ion) therapy treatment planning \cite{Grosswendt2005-ni}. Radiation damage to genes or cells begins with initial damage to DNA segments (depending on the number of relevant particle interactions) and, consequently, to particle track structures a few nanometers in size. The infrared quantum dots can be applied in nanodosimetric detectors for absolute prediction, detection and measurements of radiation damage in multiple sites of DNA lesion clusters, which is also expected to contribute to radiation protection as regular DNA damage leads towards cancer \cite{Alhmoud2020-jp}. The future of IR QDs in medicine lies in highly bright, biocompatible, and multifunctional nanoprobes that integrate deep-tissue imaging, precision therapy, and real-time monitoring. A primary research focus will be on developing non-toxic, heavy-metal-free IR QDs, as well as improving biocompatibility, clearance, and regulatory acceptance for human applications. With advances in material engineering, imaging technologies, and safety profiles, IR QDs are expected to transition from preclinical research to clinical applications within the coming decade.

\section{Future Perspectives for Space and Astrophysics}
Building on the wavelength‐multiplexed filter‐array spectrometer as proposed in \cite{Bao2015-nature14576}, IR-QDs can be composition and size-engineered so that each sub-filter exhibits a distinct absorption band across the mid and long-wave IR. After calibrating the filter–wavelength response, the resulting sensing matrix can be inverted using regularised algorithms to reconstruct the incident spectrum on a single detector or a compact focal-plane imager, extending grating-free QD spectrometers from the visible/NIR into the MWIR and well into the LWIR. Together with modern reconstruction methods for miniaturised computational spectrometers \cite{Xue2024-AdvSci-CompSpec}, IR-QD filter arrays offer a path to compact, low-SWaP spectrometers that broaden spectral coverage for planetary and lunar remote-sensing payloads.

This IR-QD spectrometer concept aligns with NASA Goddard's spacecraft-as-sensor vision, where printed QD spectrometers turn spacecraft surfaces into lightweight, distributed sensing arrays \cite{Hille2022-bg,Bao2015-nature14576}. Using IR-active QDs would extend this platform from the visible/NIR into the MWIR/LWIR, enabling in-situ thermal and chemical mapping while meeting strict size, weight, and power constraints.

The most immediate aerospace use at these wavelengths is free-space optical (FSO) communications for inter-satellite links and space-to-ground downlinks. In this case, the QD will be used as an emitter. The FSO systems commonly favour $\sim 1550$ nm carriers because of eye-safety limits and reduced solar/sky background, while still offering high data-rate, narrow-beam links; the Deep Space Optical Communications (DSOC) further underscores the viability of $1.55~\mu m$ hardware for deep-space regimes \cite{Alimi2024FSO,Vafaie2021SWIRBenefits}. QD emitters near $1.5~\mu m$ therefore align with established optics, InGaAs detectors, and silicon-photonics components, enabling low-cost, arrayed transmitters, modulated beacons, and redundant optical crosslinks \cite{Pradhan2022QLED1550}.

Beyond links, $1.5~\mu m$ QD emitters enable cooperative optical navigation (formation-flying beacons/retroreflector tracking) and are strong candidates for eye-safer, short-range lidar/altimetry in proximity operations; they also serve as compact, spectrally tailorable in-flight calibration sources for SWIR imagers and as luminescent coatings for remote thermometry of structures and thermal-protection systems \cite{Abadi2019SelfAlignFSO,Vafaie2021SWIRBenefits,EP2280255QDCal,Geiregat2023SiN,Far2023ThermoReview}.  Emerging integration of PbS-QD optoelectronics with SiN waveguides points to fully integrated on-chip emitters/detectors and calibration modules at \(\sim1.55\,\mu\)m for cubesat-class platforms \cite{Geiregat2023SiN}.

\section{Conclusion and Outlook}
In summary, infrared quantum dots are a promising and advancing technology with application prospects in optoelectronics, photovoltaics, high-energy physics, medicine, and space. Their advantages include bright, tunable fluorescence; reduced photobleaching compared to organic dyes for data transmission; suitability for chromatic tracking, advanced bioimaging, and diagnostics; fluorescence-guided surgery with real-time visualisation; integration with AI-driven imaging systems; theranostics; and targeted and personalised medicine, with enhanced biosafety regulations. Future research should focus on enhancing the scalability and stability of infrared quantum dots, advancing surface engineering, developing non-toxic, heavy-metal-free infrared quantum dots, and defect passivation, aiming to create high-performance infrared quantum dots with superior sensitivity and efficiency. Furthermore, scalable continuous-flow synthesis and standardised testing protocols can be key factors in industrial adoption, unlocking the full potential of IR QDs for next-generation imaging, sensing, medical, high-energy physics, and space technologies.

\bibliographystyle{unsrt}
\bibliography{references}
\end{document}